\documentclass[smallextended]{svjour3} 
\usepackage[english]{babel}
\usepackage[utf8]{inputenc}
\usepackage{pgf,tikz,pgfplots}
\pgfplotsset{compat=1.17}
\usetikzlibrary{arrows}
\usepackage{amssymb}
\usepackage{bm}
\usepackage{csquotes}
\expandafter\let\csname equation*\endcsname\relax
\expandafter\let\csname endequation*\endcsname\relax
\usepackage{amsmath}
\usepackage{amsfonts}
\usepackage{amstext}
\usepackage{physics}
\usepackage{color}
\usepackage{hyperref}
\usepackage{enumerate}
\usepackage{graphicx}
\usepackage{float}
\usepackage{setspace} 
\usepackage{lipsum}
\usepackage{dcolumn}
\usepackage{caption}
\usepackage{subcaption}
\usepackage{epsfig}
\usepackage{epsf}
\usepackage{epstopdf}

\usepackage{enumitem}
\usepackage{xcolor}
\usepackage{tikz}
\usepackage[margin=1in]{geometry}
\usepackage{pgfplots}
\usepackage{esint}

\usepackage{scalerel}
\usepackage{stackengine}
\usepackage{braket}
\usepackage{MnSymbol}
\usepackage{titlesec}
\usepackage{verbatim}
\usepackage{tikz}

\newcommand{\vect}[1]{\boldsymbol{#1}}
\DeclareSymbolFont{usualmathcal}{OMS}{cmsy}{m}{n}
\DeclareSymbolFontAlphabet{\mathcal}{usualmathcal}

\newcommand{\be}{\begin{equation}}
\newcommand{\ee}{\end{equation}}
\newcommand{\bea}{\begin{eqnarray}}
\newcommand{\eea}{\end{eqnarray}}

\begin{document}
\title{Blast waves in the zero temperature hard sphere gas: double scaling structure}

\author{Sahil Kumar Singh\textsuperscript{1}, Subhadip Chakraborti\textsuperscript{1}, Abhishek Dhar\textsuperscript{1} and P. L. Krapivsky\textsuperscript{2,3}}

\institute{${}^{1}$ International Centre for Theoretical Sciences, Bengaluru, India-560089 \\
	${}^{2}$ Department of Physics, Boston University, Boston, MA 02215, USA \\ ${}^{3}$ Santa Fe Institute, 1399 Hyde Park Road, Santa Fe, NM 87501, USA }

\date{Received: date / Accepted: date}

\maketitle

\begin{abstract}
We study the blast generated by sudden localized release of energy in a cold gas. Specifically, we consider one-dimensional hard-rod gas and two-dimensional hard disc gas. For this problem, the Taylor-von Neumann-Sedov (TvNS) solution of Euler equations has a self-similar form. The shock wave remains infinitely strong for the zero-temperature gas, so the solution applies indefinitely. The TvNS solution ignores dissipation, however. We show that this is erroneous in the core region which, in two dimensions, expands as $t^{2/5}$ while the shock wave propagates as $t^{1/2}$. A new self-similar solution depending on the scaling variable $r/t^{2/5}$ describes the core, while the TvNS solution describes the bulk. We demonstrate this from a numerical solution of the Navier-Stokes (NS) equations and from molecular dynamics  simulations {for a gas of hard rods in two dimensions and hard rods in one dimension. In both cases, the shock front position predicted by NS equations and by the TvNS solution agrees with that predicted by molecular dynamics simulations. However, the NS equations fail to describe the near-core form of the scaling functions.}
\end{abstract}
\tableofcontents

\section{Introduction}
\label{sec:intro}

The non-equilibrium state of a fluid is usually described by hydrodynamics. However, on a microscopic level, the individual particles follow Newton's laws of motion. How does hydrodynamics emerge from microscopic Newtonian dynamics? A rigorous derivation of hydrodynamics starting from microscopic Newtonian dynamics is lacking. Phenomenological textbook derivations involve applying conservation of mass, momentum and energy to a parcel of fluid, and there are also derivations based on the Boltzmann equation \cite{landau9}. The latter assume that the hydrodynamic limit is achieved. More precisely, the spatial variations of the coarse-grained variables are sufficiently slow, i.e., local equilibrium is achieved, and the deviations from local equilibrium are small. Thus to zeroth order, one assumes local equilibrium and gets the Euler equations which do not have dissipation. To next order, one includes deviations from local equilibrium to get Navier-Stokes equations which include dissipation as higher order derivative corrections to the Euler equations. 
\par
In some cases the hydrodynamic limit may not be achieved. For example, when there is shock formation, there is a sharp variation in the hydrodynamic fields, and one may expect hydrodynamics to break down. One situation where there can be shock formation is when the energy released from a blast is left to propagate freely in an otherwise cold gas. Also, the center of the blast is a low density region where collisions are rare and local equilibrium is difficult to reach. Thus the blast problem is a classic problem to test the validity of hydrodynamics. It was first studied in the context of atomic explosions and underlies the behavior of many astrophysical systems, see \cite{Ostriker,Goodman,Gal-Yam,Chevalier,Burrows}. An exploding star sends a blast wave into the stellar medium and later into the interstellar medium. The released energy is tremendous in astrophysical applications, and hence the assumption of a zero temperature, zero pressure ambient gas, in which the explosion propagates, is an excellent approximation.

One of the central problems is to determine how the shock wave advances and to compute the hydrodynamic fields behind the shock. The blast problem is traditionally studied in the framework of an {\em ideal} compressible gas neglecting the effect of dissipation. Dimensional analysis~\cite{sedov,landau} allows one to express the radius $R(t)$ of the shock wave through the time $t$ counted from the moment of explosion, the released energy $E_0$ and the density $\rho_{\infty}$ in front of the shock
\begin{equation}
\label{R-d}
R(t) = \left(\frac{E_0t^2}{A_d\, \rho_{\infty}}\right)^\frac{1}{d+2}.
\end{equation}
Here $d$ is the spatial dimension. The amplitude $A_d$ is a priori unknown, fixing it is a part of the solution. It was shown by Taylor, von Neumann, and Sedov~\cite{taylor1,taylor1950,vN,sedov1,sedov} that at long times the hydrodynamic fields behind the shock front have a self-similar scaling form that can be obtained from a solution of the Euler equations. This solution will be referred to as the Taylor-von Neumann-Sedov (TvNS) solution.  One of the main issues that we address here is the   effect of  dissipation on the TvNS predictions.   
\par
By writing the Navier-Stokes (NS) equations, one can easily see  that dissipation in the flow behind the shock is indeed negligible, apart from the region near the center of the explosion where the temperature obtained from ideal hydrodynamics, i.e., Euler equations, diverges. This is physically impossible since the energy injected is finite. Dissipation is thus expected to affect the flow in the region surrounding the origin which we will refer to as the core region.  Earlier studies of the influence of heat conduction~\cite{Bethe,zel,Oppenheim,Steiner} and viscosity~\cite{Johnny,Brode,Latter,Plooster} on the blast were performed in the realm of atomic explosions.  
\par

Recently, a comparison was  made between the results of Newtonian dynamics from molecular dynamics~(MD) simulations and that of hydrodynamics in the one-dimensional (1D) alternating mass hard particle (AHP) gas for a blast-like~\cite{chakraborti,ganapa}  and splash~\cite{chakraborti2022} initial condition, and it was found that the TvNS solution describes, surprisingly accurately, the results of MD simulations except in the core of the blast. As expected, due to the effect of dissipation, the TvNS solution  fails in the core. It was noted in \cite{chakraborti,ganapa} that the core has size $X(t) \sim t^{38/93}$ which, at large times,  is much smaller than the size of the blast given from Eq.~(\ref{R-d}), for $d=1,$ by $R(t)\sim t^{2/3}$. The core region does not have a sharp boundary, unlike the shock wave front. An analysis of the hydrodynamic equations including dissipative terms yields a  different scaling solution  in the core which  was compared with results from MD. While dissipation kills the unphysical divergences of the Euler solution, the detailed agreement between simulations and dissipative hydrodynamics  is not very good in the core region. One reason could be the presence of anomalous heat transport in $d=1$~\cite{LLP2003,dhar2008}.  Fourier's law in one dimension breaks down, and it is replaced by a non-local law \cite{kundu1,kundu2}. Thus, we expect hydrodynamics to break down in one dimension, and such a breakdown has indeed been observed in \cite{hurtado}. Thus, the near-perfect agreement found between the TvNS solution and molecular dynamics simulations in \cite{chakraborti,ganapa} is quite surprising. It can only be speculated that since the dissipative region is small in the 1D blast problem, much of the perturbed region in the gas is described by dissipation-less Euler equations.

A natural question is to investigate the effect of dissipation in higher dimensions. The first study \cite{Trizac2} comparing the predictions of molecular dynamics with those of Euler equations in two dimensions (2D) found a reasonable agreement between the two.  Other studies in two and three-dimensional hard-sphere gases~\cite{joy1,joy2,amit}   found  disagreement between the TvNS prediction and MD simulations (even away from the core region).
 It turns out~\cite{private}  that the reason for the observed discrepancy in \cite{joy1} arose from an incorrect estimation of the value of initial energy of the blast used in the simulation. We have verified this in our MD simulations where we find perfect agreement with TvNS in the outer region~(see Fig.~\eqref{fig3} .

In this work we first consider the blast in a 2D gas of hard discs and in a toy  2D gas of point particles. We also  consider  a 1D gas of hard rods with alternating masses (to ensure non-integrability) to see if the finite rod size changes the remarkable agreement between MD results and hydrodynamics seen in \cite{chakraborti,ganapa}. Our main results are:

(i) For the 2D system, we show that, as in 1D, the dissipative terms in the NS equations introduce a new growing length scale which we denote by $X(t)$.  We find a new self-similar solution in the core, different from the TvNS scaling solution. This core scaling solution is verified from the  numerical solution of the NS equations. Thus there are two scaling regions: the bulk region described by TvNS scaling and the core region described by the dissipative scaling. We further show that the two regions are connected by the rules of asymptotic matching.

(ii)  We find that NS results agree perfectly with TvNS in the bulk region and {also agree with the  MD simulation results. The MD results satisfy TvNS scaling in the bulk region and also the core scaling in the central region. However,  the scaling functions in the core  region are different from the predictions of hydrodynamics. }

(iii) For the gas of hard rods in 1D we find an excellent agreement between TvNS, NS and MD simulations in the bulk region, including for the position of the shock front and for the scaling functions. In the blast core, the NS scaling is observed in the MD simulations, but the scaling functions do not agree with those obtained from NS.

\par
We summarize here the plan of the paper.  In Sec.~(\ref{sec:2.1}), we review the TvNS solution. In Sec.~(\ref{sec:2.2}), we discuss the effect of dissipation and the new scaling solution that exists in the core. In Sec.~(\ref{sec:2.3}), we present numerical results of the NS equations which verify the core scaling forms. We also show that MD simulations  satisfy this scaling and point out the {agreement} between MD and NS.  In Sec.~(\ref{sec:3}), we present results on the 2D gas when virial corrections in the equation of state are ignored.   In Sec.~(\ref{sec:4}), we present results on hard rods. Sec.~\eqref{sec:ideal}  contain a brief discussion on the form of the inner and outer solutions in arbitrary dimensions. We conclude in Sec.~(\ref{sec:discussion}) with a discussion. 

\section{Blast in the 2D hard-disc gas}
\label{sec:2}
We first consider a gas of hard discs of diameter $a$, and mass $m$, whose only interaction is via elastic collisions  that conserve momentum and energy. Particles move with constant velocities between collisions. The finite size of the particles implies that the equation of state is no longer given by the ideal gas equation. This feature complicates the solution of the hydrodynamic equations. At a microscopic level, the blast initial condition consists of taking an infinite gas of discs that are distributed uniformly in the 2D plane with mass density,  $\rho_\infty$, and all at rest.  Particles inside a localized region are then excited such that their total energy is $E_0$, and the total momentum is zero. The excitation then evolves in a radially symmetric way and we are interested, in particular, in how  the three conserved fields  of mass density, $\rho(r)$, momentum density, $p(r)$, and energy density, $e(r)$,  evolve with time. We first review the TvNS solution as presented in \cite{joy1}, and then we discuss the effect of dissipation by considering the NS equations.

\subsection{TvNS solution}
\label{sec:2.1}
The classic TvNS solution was first obtained for a 3D ideal gas using Euler equations and ignoring any virial corrections caused by the finite cross sectional area of the colliding molecules. A similar analysis was done in \cite{joy1} for hard-disc gas, but by including the virial corrections in the equation of state. 
\par
The first step in finding TvNS solution is to find how the shock front $R(t)$ grows as a function of time. As discussed above this can  be found from purely dimensional analysis. The only variables that the shock front can depend on at large times are time $t$, energy injected $E_0$, and the ambient density $\rho_{\infty}$. Applying dimensional analysis, we get 
    $R(t)=[E_0t^2/(A_d\rho_{\infty})]^{\frac{1}{d+2}}$,  where $A_d$ is an unknown dimensionless constant,  that only depends on the volume fraction $(\rho_\infty/m) V_d a^{d}$, where $V_d=2^{-d}\pi^{d/2}/\Gamma(d/2+1)$ is the volume of a  $d$-dimensional sphere of unit diameter. 
The Euler equations in any dimension are given by:
\begin{subequations}
\label{euler-dD}
\begin{gather}
    \partial_t\rho+\partial_i(\rho v_i)=0,\\
    \partial_t(\rho v_i)+\partial_j(\rho v_iv_j)+\partial_i P=0,\\
    \partial_t(\rho e)+\partial_i(\rho ev_i)+\partial_i(Pv_i)=0,
    \end{gather}
\end{subequations}
where $P$ is the pressure, which can be related to the other hydrodynamic fields via equation of state, and $e$ is the energy per unit mass, $e={v^2}/{2}+{dT}/{2}$.  Henceforth, in most of the discussions here we set $a=k_B=m=1$, unless otherwise specified.

For the 2D hard disc gas, the equation of state at low densities is given by the virial equation of state:
\begin{equation}
    P=\rho T\Big(1+\sum_{n=2}^{\infty}B_n\rho^{n-1}\Big).
\end{equation}
Following \cite{joy1}, we only take terms up to $n=10$. The values of the constants $B_n$ are given in \cite{joy1}. According to TvNS, the blast wave initial condition evolves, at long times to a self-similar scaling form.   We define the following scaling variable:
\begin{equation}
    \xi=\frac{r}{R(t)}.
\end{equation}
In terms of this dimensionless variable, one again  uses dimensional analysis to make the following ansatz for the long time solution of Eqs.~(\ref{euler-dD}) for the density, velocity and temperature fields:
\begin{subequations}
\begin{align}
\rho&=\rho_{\infty}G(\xi),\\
v&=\frac{r}{t}V(\xi),\\
T&=\frac{r^2}{t^2}Z(\xi).
\end{align}
\label{eq:TvNSscal}
\end{subequations}
Putting these scaling forms into the Euler equations, we get the following ODEs for the scaling functions:
\begin{subequations}
\label{eq:14}
\begin{gather}
  \xi  \Big(V-\frac{1}{2}\Big)G\frac{dV}{d\xi}+\xi\frac{d}{d\xi}\left[ZG B(G)\right]-GV+GV^2+2GZ B(G)=0,\\
   \xi  \Big(V-\frac{1}{2}\Big)\frac{dG}{d\xi}+\xi G\frac{dV}{d\xi}+2GV=0,\\
    - \frac{\xi B(G)}{G}\frac{dG}{d\xi}+\frac{\xi}{Z}\frac{dZ}{d\xi}+2\frac{(V-1)}{(V-1/2)}=0, 
     \end{gather}
\end{subequations}
where
\begin{equation}
     B(G)=1+\sum_{n=2}^{\infty}B_n\rho_{\infty}^{n-1}G^{n-1}.
\end{equation}
Under the dynamics of the Euler equations, energy is conserved, and this gives us the further constraint:
\begin{equation}
    \int_0^{R(t)} dr~2\pi r \rho e=E_0,
\end{equation}
which, in terms of the scaling functions becomes:
\begin{equation}
    2\pi\int_0^{1} d\xi~\xi^3 G(\xi)\Big(\frac{V^2(\xi)}{2}+Z(\xi)\Big)=A_2.
    \label{eq:17}
\end{equation}
The boundary conditions required to solve these ODEs are provided by the Rankine-Hugoniot conditions which specify the discontinuity of the fields across the shock: 
\begin{subequations}
\begin{gather}
    \frac{\rho(R)v(R)}{\rho(R)-\rho_{\infty}}=\dot{R},\\
    \frac{\rho(R) v^2(R)+P(R)}{\rho(R)v(R)}=\dot{R}, \\
     \frac{\rho(R) v(R) e(R) +P(R) v(R)}{\rho(R)e(R)}=\dot{R}.
    \end{gather}
\end{subequations}
These, in terms of the scaling functions are given by:
\begin{subequations}
\begin{gather}
    \frac{1}{G(1)}\left(1+\frac{2}{  B[G(1)]}\right)=1,\\
    V(1)=\frac{1}{G(1)~B[G(1)]},\\
    Z(1)=\frac{1}{2}V^2(1).
    \end{gather}
\end{subequations}
These equations can be solved numerically to find the values of the scaling functions at the shock front, which can be used as boundary values to solve the ODEs. The dimensionless constant $A_2$ can be found using Eq.~(\ref{eq:17}).
\subsection{Effect of dissipation on scaling in the blast core}
\label{sec:2.2}
In the Euler framework, it is assumed that the hydrodynamic fields satisfy the Euler equations throughout the blast and that dissipative terms can be neglected. This assumption is valid only if the dissipative terms corresponding to the TvNS solution are small compared to the Euler terms. The TvNS solution predicts a divergence $Z(\xi) \sim \xi^{-4}$ for small $\xi$ which means that the temperature field will have a steep spatial variation. Since the spatial derivatives in the dissipative terms in the NS equations are of one order higher than the Euler terms, this signals that the dissipative terms corresponding to the TvNS solution are much larger than the Euler terms, thus suggesting that dissipation becomes important near the core. However, far from the core, the slopes of the various scaling functions are not steep, so dissipation is not important in those regions. Since dissipation terms far from the core are not important, we expect the ODEs to still hold in the bulk region. 
\par
We can estimate the size of the core where dissipation becomes important. The Navier-Stokes equations in $d$-dimensions are given by:
\begin{subequations}
\begin{gather}
    \partial_t\rho+\partial_i(\rho v_i)=0,\\
     \partial_t(\rho v_i)+\partial_j(\rho v_iv_j)+\partial_j\sigma_{ij}=0,\\
     \partial_t(\rho e)+\partial_i(\rho ev_i)+\partial_i(\sigma_{ij}v_j)+\partial_i(-\kappa\partial_i T)=0,
    \end{gather}
\end{subequations}
with the stress tensor $\sigma$ given by:
\begin{equation}
    \sigma_{ij}=P\delta_{ij}-\mu[\partial_iv_j+\partial_jv_i-\frac{2}{d}(\nabla\cdot\vect{v})\delta_{ij}]-\zeta(\nabla\cdot\vect{v})\delta_{ij},
\end{equation}
where $\mu$ is the shear viscosity, $\zeta$ is the bulk viscosity and $\kappa$ is the heat conductivity.
Kinetic theory predicts that $\mu=D_{\mu}\sqrt{T}$, $\kappa=D_{\kappa}\sqrt{T}$ and $\zeta=0$ for a monoatomic gas in the low density limit \cite{landau9,beijeren}, where $D_{\kappa}$ and $D_{\mu}$ are some constants. However, we emphasize that these results are only first order terms in a virial expansion \cite{landau9}. For a radially symmetric blast, all the hydrodynamic fields are radially symmetric. Using this fact, the NS equations become (for $d=2$):
\begin{subequations}
    \label{eq:NS2d}
\begin{gather}
    \partial_t\rho+\frac{\rho v}{r}+\partial_r(\rho v)=0,\\
    \rho(\partial_tv+ v\partial_rv)+\partial_rP=\frac{1}{r^2}\partial_r[r^2\mu(\partial_rv-r^{-1}v)], \label{momeq}\\
    \rho(\partial_te+ v\partial_re)+\frac{1}{r}\partial_r(rPv)=\mu(\partial_rv-\frac{v}{r} )^2+\frac{v}{r^2}\partial_r[r^2\mu(\partial_rv-\frac{v}{r})]+\frac{1}{r}\partial_r(r\kappa\partial_rT).
    \end{gather}
\end{subequations}
It is known~\cite{joy1},  that the TvNS solution predicts (in 2D) that for small $\xi$,  $G\sim\xi^2$, $V-1/4 \sim \xi^4$ and $Z\sim\xi^{-4}$ [see also Sec.~(\ref{sec:3})]. Using this small $\xi$ behaviour predicted by TvNS solution, we can estimate if the dissipation terms are indeed small, e.g, by looking at the ratio of the terms on the right and left hand sides of Eq.~(\ref{momeq}). This shows that dissipation is small  for $r\gtrsim {D_{\mu}^{1/5}E_0^{1/5}}{\rho_{\infty}^{-2/5}}t^{2/5}$, thus verifying the assumption leading to the TvNS solution. However, for $r\lesssim {D_{\mu}^{1/5}E_0^{1/5}}{\rho_{\infty}^{-2/5}}t^{2/5}$, dissipation terms are larger than Euler terms, and thus the assumption leading to TvNS solution breaks down. Hence, we expect TvNS solution not to be valid for $r\lesssim {D_{\mu}^{1/5}E_0^{1/5}}{\rho_{\infty}^{-2/5}}t^{2/5}$. This introduces a new length scale, which is an estimate of the size of the core, 
\begin{equation}
X(t) = \frac{D_{\mu}^{1/5}E_0^{1/5}}{\rho_{\infty}^{2/5}} t^{2/5}, \label{core-l}
\end{equation} 
and which grows with time as $t^{2/5}$. From this length scale, we can find an estimate for the density, velocity and temperature in the core, which we denote respectively by $\rho^*$, $v^*$ and $T^*$. This will give us new density, velocity and temperature scales (here $\xi_*={X(t)}/{R(t)}$ is the scaled size of the core):
\begin{subequations}
\begin{gather}
    \rho^*= \rho_{\infty}\xi_*^2= b_{\rho}t^{-1/5},\\
    v^*=\frac{X(t)}{t}= b_{v} t^{-3/5},\\
     T^*=\frac{X^2(t)}{t^2}\xi_*^{-4}\sim b_Tt^{-4/5},
     \end{gather}
\end{subequations}
where  $b_{\rho}={A^{1/2} D_{\mu}^{2/5}\rho_{\infty}^{7/10}}{E_0^{-1/10}}$, $b_v=D_{\mu}^{1/5} {\rho_{\infty}^{-2/5}} E_0^{1/5}$ and $b_T={A^{-1}D_{\mu}^{-2/5}\rho_{\infty}^{-1/5}}{E_0^{3/5}}$. From these scales, we can construct a new scaling form for the various hydrodynamic fields:
\begin{subequations}
\label{eq:35Cscal}
\begin{gather}
    \rho=t^{-\frac{1}{5}}\tilde{G}(\eta),\\
    v=t^{-\frac{3}{5}}\tilde{V}(\eta),\\
    T=t^{-\frac{4}{5}}\tilde{Z}(\eta),
    \end{gather}
\end{subequations}
where we have defined the new scaling variable 
\begin{equation}
\eta=\frac{r}{t^{2/5}},
\end{equation}
and the constants have been absorbed into the scaling functions $\tilde{G}$, $\tilde{V}$ and $\tilde{Z}$.
\par

To get the ODEs satisfied by the core scaling functions $\tilde{G}(\eta)$, $\tilde{V}(\eta)$ and $\tilde{Z}(\eta)$, we plug the scaling ansatz (Eqs.~(\ref{eq:35Cscal})) into the NS equations and take the long time limit. We get the following coupled ODEs:
\begin{subequations}
\label{eq:38}
\begin{gather}
\label{eq:38a}
    -\frac{\tilde{G}}{5}-\frac{2\eta \tilde{G}'}{5}+\frac{(\eta \tilde{G}\tilde{V})'}{\eta}=0,\\
    \label{eq:38b}
    (\tilde{Z}\tilde{G})'=0,\\
    \label{eq:38c}
     -\frac{4}{5}\tilde{Z}\tilde{G}-\frac{2\eta \tilde{G}\tilde{Z}'}{5}+\tilde{G}\tilde{V}\tilde{Z}'+\frac{\tilde{Z}\tilde{G}\tilde{V}}{\eta}+(\tilde{G}\tilde{V}\tilde{Z})'=\frac{D_{\kappa}\tilde{Z}^{1/2}\tilde{Z}'}{\eta}+\frac{2}{3}(D_{\kappa}\tilde{Z}^{3/2})''.
    \end{gather}
\end{subequations}
We note the following: (a) The virial corrections do not appear in these equations because they vanish in the long time limit; (b) only the thermal conductivity term makes an appearance [this is true also if we consider the form of transport coefficients in Eq.~(\ref{eq:EC})].  These ODEs are first order in $\tilde{G}$ and $\tilde{V}$ and second order in $\tilde{Z}$. Thus, we need 4 boundary values: two for each $\tilde{G}$ and $\tilde{V}$, and two for $\tilde{Z}$. Two of the boundary conditions can be determined using the radial symmetry of the problem. Because of radial symmetry, there is no flow velocity and heat flow in the origin. This gives $\tilde{V}(\eta=0)=0$ and $\tilde{Z}'(\eta=0)=0$. The rest two boundary conditions can be determined from the asymptotic matching condition, which is given by:
\begin{equation}
    h(\eta\rightarrow\infty)=h(\xi\rightarrow 0),
    \label{eq:matching}
\end{equation}
where $h$ is some hydrodynamic field.
\par
We call the solution obtained from the  ODEs in Eqs.~(\ref{eq:14}) for the TvNS scaling functions as the {\it{outer solution}}, and those obtained from the ODEs in Eqs.~(\ref{eq:38}) for the core scaling functions as the {\it{inner solution}}. We have not succeeded in finding analytical expressions for the inner solution. However, approximate solutions can be found in the small $\eta$ region,  as we now show.
Since $\tilde{V}(0)=0$ and $\tilde{Z}'(0)=0$ we can take for small $\eta$:
\begin{equation}
    \tilde{V}(\eta)\approx V_0\eta^{\alpha}\text{,  $\alpha>0$},
\end{equation}
\begin{equation}
    \tilde{Z}'(\eta)\approx F_0\eta^{\beta}\text{,   $\beta>0$}.
\end{equation}
Thus,
\begin{equation}
    \tilde{Z}(\eta)=F_0\frac{\eta^{1+\beta}}{1+\beta}+C_1,\text{  for small $\eta$}.
\end{equation}
Using these and Eq.~(\ref{eq:38b}), we have
\begin{equation}
    \tilde{G}=\frac{C_2}{\tilde{Z}}\approx\frac{C_2}{C_1}\Big(1-\frac{F_0\eta^{1+\beta}}{C_1(1+\beta)}\Big),\text{ for small $\eta$}.
\end{equation}
Putting these into Eq.~(\ref{eq:38a}) and comparing powers and taking small $\eta$ limit, we get $\alpha=1$, $V_0=1/10$. Using these values for $\alpha$ and $V_0$ and Eq.~(\ref{eq:38c}), we get:
\begin{figure}
\centering
\begin{subfigure}{.33\textwidth}
\centering
\includegraphics[width=1.0\linewidth]{{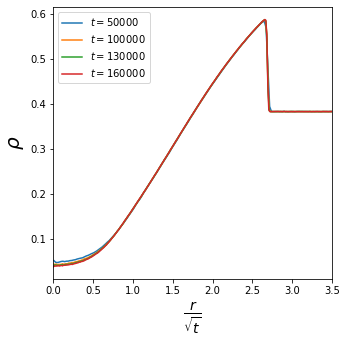}}
\caption{}
\end{subfigure}%
\begin{subfigure}{.33\textwidth}
\centering
\includegraphics[width=1.0\linewidth]{{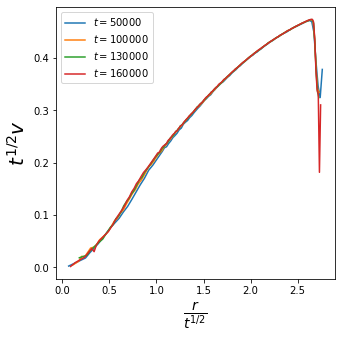}}
\caption{}
\end{subfigure}%
\begin{subfigure}{.33\textwidth}
\centering
\includegraphics[width=1.0\linewidth]{{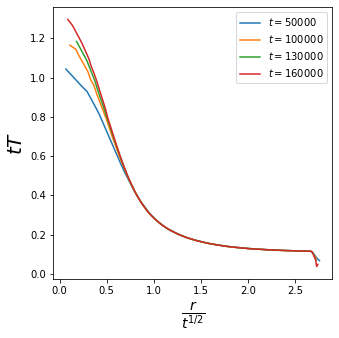}}
\caption{}
\end{subfigure}
\begin{subfigure}{.33\textwidth}
\centering
\includegraphics[width=1.0\linewidth]{{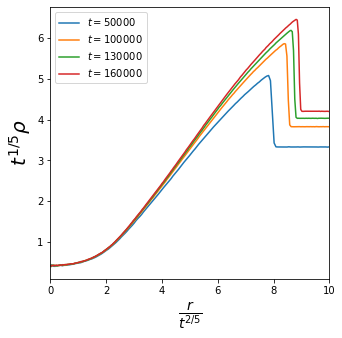}}
\caption{}
\end{subfigure}%
\begin{subfigure}{.33\textwidth}
\centering
\includegraphics[width=1.0\linewidth]{{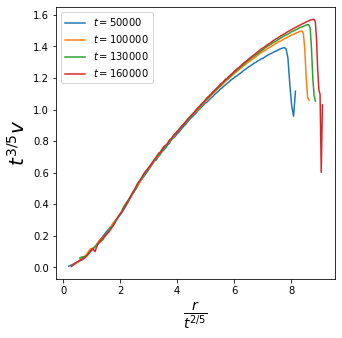}}
\caption{}
\end{subfigure}%
\begin{subfigure}{.33\textwidth}
\centering
\includegraphics[width=1.0\linewidth]{{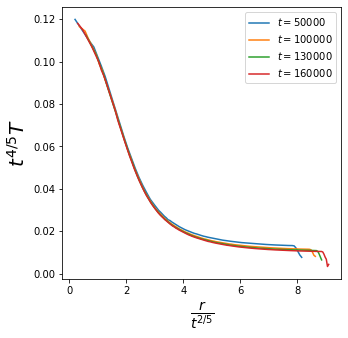}}
\caption{}
\end{subfigure}
\caption{{\bf MD simulations of hard discs}: The three hydrodynamic fields, obtained from MD simulation data of Ref.~\cite{joy1}, plotted in (a,b,c) according to the expected TvNS scaling form  [Eqs.~(\ref{eq:TvNSscal})] and in (d,e,f) according to the expected core scaling form [Eqs.~(\ref{eq:35Cscal})]. We can see that in (a,b,c)  there is a very good collapse of data  everywhere except near the core, where the scaling is not very good, especially for the temperature field.  On the other hand in (d,e,f) we see excellent data collapse near the core. Thus MD simulations exhibit the expected core and bulk scaling forms.\label{fig18}}
\end{figure}
\begin{eqnarray}
    -\frac{3C_2}{5}-\frac{2\eta C_2F_0}{5C_1}\Big(\eta^{\beta}-\frac{F_0\eta^{1+2\beta}}{C_1(1+\beta)}\Big)+\frac{C_2F_0}{10C_1}\Big(\eta^{1+\beta}-\frac{F_0\eta^{2+2\beta}}{C_1(1+\beta)}\Big)\nonumber\\
    -F_0D_{\kappa}\sqrt{C_1}\Big(\eta^{\beta-1}+\frac{F_0\eta^{2\beta}}{2C_1(1+\beta)}\Big)-D_{\kappa}\sqrt{C_1}F_0\beta\eta^{\beta-1}=0.
\end{eqnarray}
Since $\beta>0$, the equation above can only be satisfied when $\beta=1$ and
\begin{equation}
    F_0=-\frac{3C_2}{10D_{\kappa}\sqrt{C_1}}.
\end{equation}
The only unknown constants are $C_1$ and $C_2$, which can be fixed from boundary values determined from either asymptotic matching or the data from the numerical solution of the full NS  Eqs.~(\ref{eq:NS2d}).
\par
\begin{figure}
\centering
\begin{subfigure}{.35\textwidth}
\centering
\includegraphics[width=1.0\linewidth]{{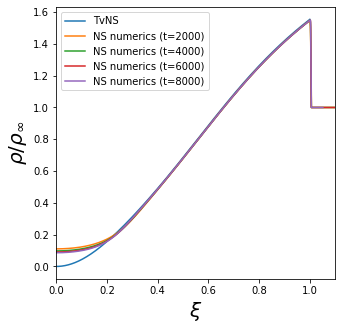}}
\caption{}
\end{subfigure}%
\begin{subfigure}{.35\textwidth}
\centering
\includegraphics[width=1.0\linewidth]{{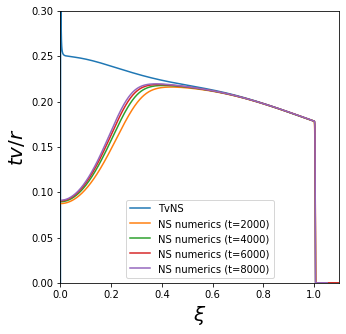}}
\caption{}
\end{subfigure}%
\begin{subfigure}{.35\textwidth}
\centering
\includegraphics[width=1.0\linewidth]{{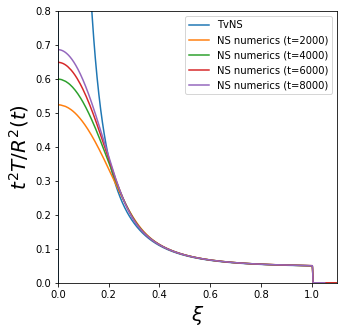}}
\caption{}
\end{subfigure}
\begin{subfigure}{.35\textwidth}
\centering
\includegraphics[width=1.0\linewidth]{{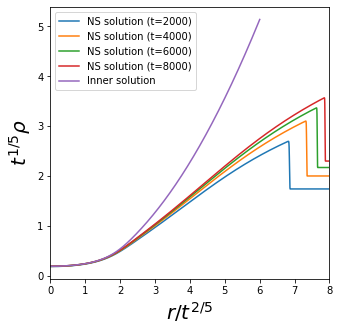}}
\caption{}
\end{subfigure}%
\begin{subfigure}{.35\textwidth}
\centering
\includegraphics[width=1.0\linewidth]{{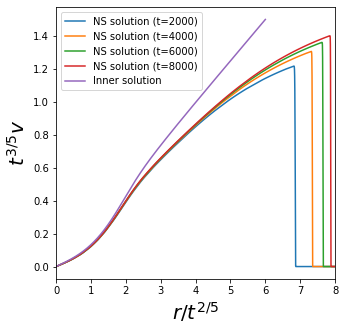}}
\caption{}
\end{subfigure}%
\begin{subfigure}{.35\textwidth}
\centering
\includegraphics[width=1.0\linewidth]{{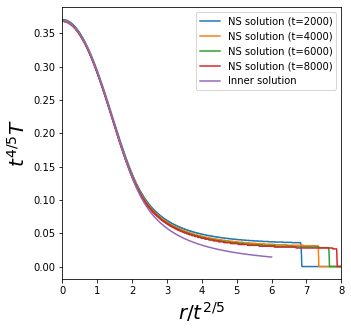}}
\caption{}
\end{subfigure}
\caption{{\bf NS solution for hard discs}:    The three hydrodynamic fields, obtained from numerical solution of NS equations in Eq.~(\ref{eq:NS2d}), are plotted in (a,b,c) according to the expected TvNS scaling form  [Eqs.~(\ref{eq:TvNSscal})] and in (d,e,f) according to the expected core scaling form [Eqs.~(\ref{eq:35Cscal})]. We can see that in (a,b,c) there is a very good collapse of data  everywhere except near the core. On the other hand in (d,e,f) we see excellent data collapse near the core. Thus the NS solutions exhibit the expected core and bulk scaling forms. In (a,b,c) we also see a good fit of the scaled data to the TvNS scaling functions, $(G,V,Z)$, also obtained from a numerical solution of Eqs.~(\ref{eq:14}).
In (d,e,f) we see a reasonable fit in the core region to the core scaling functions,  $(\tilde{G},\tilde{V},\tilde{Z}$), obtained from a numerical solution of Eqs.~(\ref{eq:38}) with boundary conditions obtained from the solution of the NS equations in  Eqs.~(\ref{eq:NS2d}). 
The parameter values were taken as $E_0=4.0$, $\rho_{\infty}=0.382$,  and the Henderson equation of state was used. \label{fig1}}
\end{figure}

Thus to first non-zero order in $\eta$, we have:
\begin{subequations}
\begin{gather}
    \tilde{G}(\eta)=\frac{C_2}{C_1}\Big(1+\frac{3C_2\eta^{2}}{20D_{\kappa}C_1^{3/2}}+\ldots\Big),\\
    \tilde{V}(\eta)=\frac{\eta}{10}+\ldots,\\
    \tilde{Z}(\eta)=C_1-\frac{3C_2\eta^{2}}{20D_{\kappa}\sqrt{C_1}}+\ldots.
    \end{gather}
\end{subequations}
\par
\subsection{Numerical results}
\label{sec:2.3}
From the discussions in the previous two subsections, we expect two different scaling form for the hydrodynamic fields, namely the  TvNS scaling form given by Eqs.~(\ref{eq:TvNSscal}) in the bulk region $X(t) \lesssim r < R(t)$, and  the core scaling form given by Eqs.~(\ref{eq:35Cscal}) in the region  $0<r\lesssim X(t)$). We now check if these predictions for the bulk and core scaling forms can be verified in data obtained from MD simulations and from numerical solutions of the NS equations. 

We use the MD simulation results from  Ref.~\cite{joy1}. They considered a gas of hard discs with diameter $a=1$, $E_0=2.0$ and $\rho_{\infty}=0.382$. The initial condition was chosen with a Gaussian temperature profile, $v(\vect{r},t=0)=0$ and $\rho(\vect{r},t=0)=\rho_{\infty}$. The parameters of the Gaussian temperature profile were chosen such that the initial total energy is $E_0$. We note here that the long-time scaling solution does not depend on the width of the Gaussian chosen as long as it corresponds to energy $E_0$. For the numerical solution of the NS equations in Eqs.~(\ref{eq:NS2d}), we considered  initial conditions that correspond to the ones used in the MD simulations. In all our computations involving hard-disc gas, we use the Henderson equation of state (EOS) \cite{henderson} instead of the truncated virial expansion for the ease of computation. The Henderson EOS is given by:
\begin{align}
    p=\rho T\frac{128+\pi^2\rho^2a^4}{8(4-\pi\rho a^2)^2}.
\end{align}
The MacCormack method \cite{maccormack} was used to solve the PDEs, with discretization given by $dx=0.05$, $dt=10^{-5}$. The initial temperature profile was taken to be a Gaussian: $T(r,0)=({T_0}/{\sqrt{2\pi\sigma^2}}) e^{-r^2/2\sigma^2}$, with $T_0={E_0}/{(\sqrt{2\pi}\rho_{\infty}\sigma)}$ and $\sigma=0.5$.
  We note that the lowest order non-zero term of the Enskog expansion for the transport coefficients  gives~\cite{gass1971,garcia2006,beijeren,kremer2010} 
\begin{align}
\label{eq:EC}
    \kappa=\frac{2}{a{\pi}^{1/2}}~T^{1/2},~~~\mu=\frac{1}{2 a {\pi}^{1/2}}~ T^{1/2},~~~\zeta=\frac{8\phi^2}{\pi^{3/2}}    T^{1/2}= \frac{\pi^{1/2} a^3}{2} ~\rho^2 T^{1/2},
\end{align}
where $\phi=\pi a^2 \rho/4$ is the volume fraction. 
 An important general property of monoatomic gases is the vanishing of the bulk viscosity in the dilute regime, $\phi\to 0$. This, together with relation $\gamma=1+2/d$, is the crucial input coming from kinetic theory \cite{landau9,beijeren,kremer2010}; phenomenological hydrodynamics is compatible with arbitrary $\gamma>1$ and $\zeta\geq 0$. When $\phi>0$, the bulk viscosity becomes positive, and Eq.~(\ref{eq:EC}) gives its leading asymptotic for small $\phi$. We note here that the  $\rho$-dependence of the bulk viscosity will not change the core scaling, as the core is a low density region where bulk viscosity is negligible.
 
 In our NS numerics we 
  followed  Ref.~\cite{amit} and chose  (with $a=1$)  the values  $\kappa=\mu={(\sqrt{\pi}}/{8}) \sqrt{T}$ and $\zeta=0$, which are  
 different from those  in Eqs.~(\ref{eq:EC}). 
We have verified that our main conclusions remain unchanged on using Eqs.~(\ref{eq:EC}) for the transport coefficients. 

\begin{figure}
\centering
\begin{subfigure}{.35\textwidth}
\centering
\includegraphics[width=1.0\linewidth]{{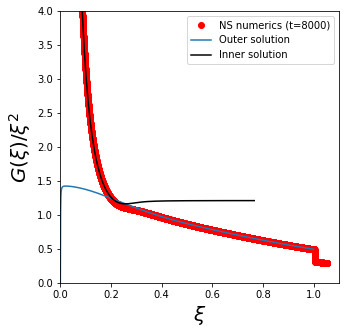}}
\caption{}
\end{subfigure}%
\begin{subfigure}{.35\textwidth}
\centering
\includegraphics[width=1.0\linewidth]{{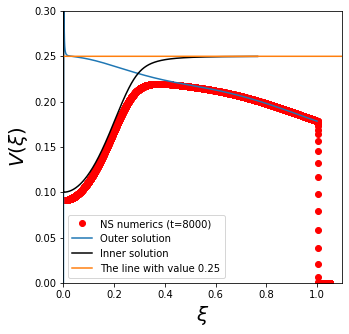}}
\caption{}
\end{subfigure}%
\begin{subfigure}{.35\textwidth}
\centering
\includegraphics[width=1.0\linewidth]{{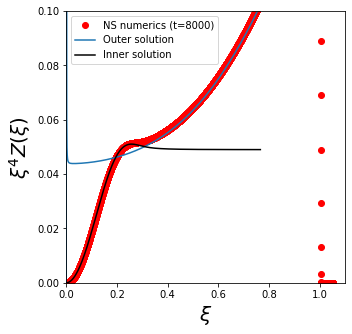}}
\caption{}
\end{subfigure}%
\caption{{\bf Asymptotic matching in hard discs}: Figure comparing the inner solution and outer solution with the full solution of NS equations. The boundary values for finding the inner solution were taken from the full solution of the NS equations. Also shown in the middle plot is the line with height $0.25$ which is the near core value of $V(\xi)$ predicted by the TvNS solution.  We see that the near core asymptotic of the outer solution matches with the far core asymptotic of the inner solution, which is nothing but the rule of asymptotic matching.\label{fig4}}
\end{figure}

{\bf MD simulations}:
In Fig.~(\ref{fig18})  we show the results of  MD simulation data and check if they satisfy the predicted scaling. In Figs.~(\ref{fig18}a,\ref{fig18}b,\ref{fig18}c), following Eqs.~(\ref{eq:TvNSscal}), we plot the scaled fields $(\rho,t^{1/2} v, t T)$  as a function of $r/t^{1/2}$, at different times, and  find that this  gives  a very good data collapse in the region  far from the core. However, as expected, the data collapse near the core is not quite good. In Figs.~(\ref{fig18}d,\ref{fig18}e,\ref{fig18}f),  we follow Eqs.~(\ref{eq:35Cscal}) and plot the scaled fields $(t^{1/5}\rho,t^{3/5} v, t^{4/5} T)$ as a function of $r/t^{2/5}$ and see that this scaling  gives a much better data collapse in this region. This establishes that there are two different regions with different scaling.

{\bf NS equations}: We now discuss the results obtained from the numerical solution of NS equations.  In Fig.~(\ref{fig1}), we plot the  hydrodynamic fields obtained from the numerical solution of the NS equations. In Figs.~(\ref{fig1}a,\ref{fig1}b,\ref{fig1}c) we verify that these satisfy the TvNS scaling form in the bulk region and agree very well with the TvNS scaling functions. However, in the core we see a clear departure from TvNS scaling. In Figs.~(\ref{fig1}d,\ref{fig1}e,\ref{fig1}f) we see that the core scaling form is satisfied in the core region.  We also  plot the solution of Eqs.~(\ref{eq:38}) and we see that they agree with the numerical solution of NS equations in the core region. 

{\bf Asymptotic matching}: In Fig.~(\ref{fig4}), we verify the rules of asymptotic matching by using the inner and outer solutions.  Once we have the inner solution $\tilde{G}$, $\tilde{V}$ and $\tilde{Z}$, we can find the corresponding hydrodynamic fields for density, velocity and temperature.   We can then compare how they match with the outer solution. In Fig.~(\ref{fig4}), we show this comparison and find that the $\xi\rightarrow 0$ limit of the outer solution has the same asymptotic as the $\xi\rightarrow\infty$ of the inner solution. Thus we verify the rule of asymptotic matching.  
\par
\begin{figure}
\centering
\begin{subfigure}{.35\textwidth}
\centering
\includegraphics[width=1.0\linewidth]{{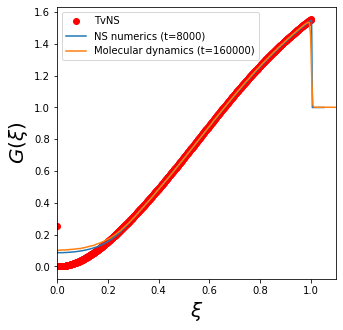}}
\caption{}
\end{subfigure}%
\begin{subfigure}{.35\textwidth}
\centering
\includegraphics[width=1.0\linewidth]{{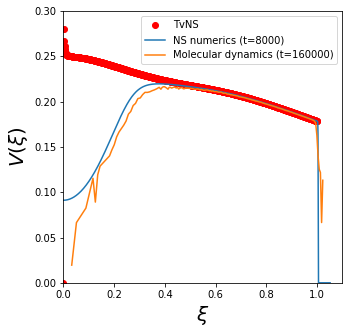}}
\caption{}
\end{subfigure}%
\begin{subfigure}{.35\textwidth}
\centering
\includegraphics[width=1.0\linewidth]{{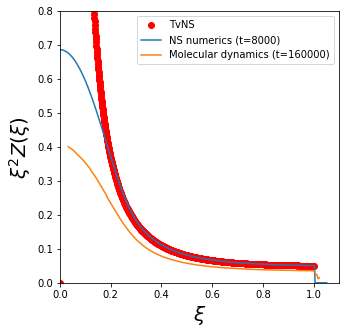}}
\caption{}
\end{subfigure}
\caption{{\bf 2D hard discs}: {Comparison of the NS solution and TvNS with MD simulations.  We  see that the MD data matches with both the NS or TvNS solution in the bulk region.} \label{fig3}}
\end{figure}

Finally in Fig.~(\ref{fig3}), we compare MD data of \cite{joy1}  (with the same value of $E_0$) with NS solution and TvNS solution. We see that NS and TvNS solutions agree with MD data.  The only discrepancies are in the near core behavior for the scaling functions.
\par

\section{Blast in 2D ideal gas}
\label{sec:3}
In this section, we consider the hydrodynamics of a toy model of a 2D gas, where we ignore the virial corrections in the equation of state, and take it to be that of an ideal gas. This amounts to considering point particles which in two dimensions  would mean a non-interacting gas. So naively we would not expect any evolution. However, one can work in the Boltzmann-Grad limit, that is, $a \to 0$ and $\rho \to \infty$ keeping $\rho a$ finite (in 2D). In this situation, one  expects finite transport coefficients predicted by kinetic theory, while still preserving the ideal gas equation of state (since the volume fraction $\pi \rho a^2/4 \to 0$). This system would be very difficult to simulate but we can still analyze the NS equations. One advantage of this toy model is that we can find an exact TvNS solution.

\begin{figure}
\centering
\begin{subfigure}{.35\textwidth}
\centering
\includegraphics[width=1.0\linewidth]{{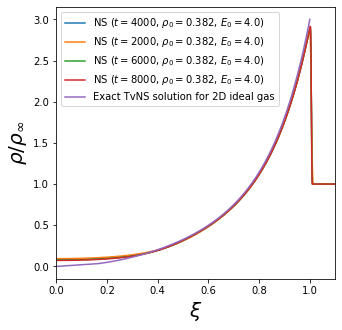}}
\caption{}
\end{subfigure}%
\begin{subfigure}{.35\textwidth}
\centering
\includegraphics[width=1.0\linewidth]{{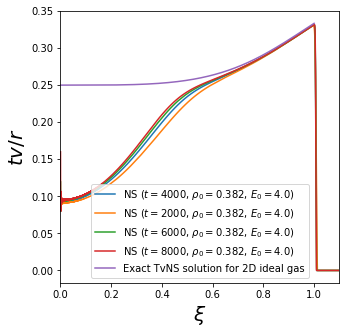}}
\caption{}
\end{subfigure}%
\begin{subfigure}{.35\textwidth}
\centering
\includegraphics[width=1.0\linewidth]{{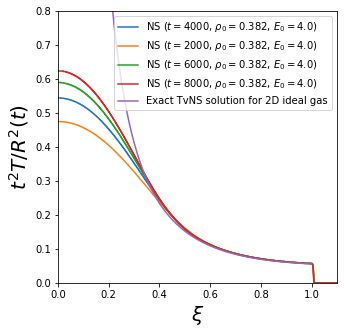}}
\caption{}
\end{subfigure}
\begin{subfigure}{.35\textwidth}
\centering
\includegraphics[width=1.0\linewidth]{{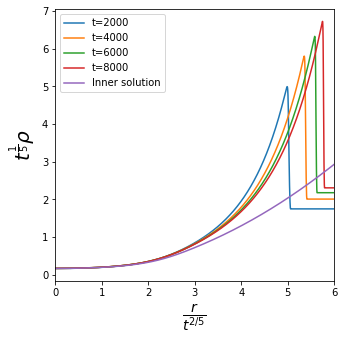}}
\caption{}
\end{subfigure}%
\begin{subfigure}{.35\textwidth}
\centering
\includegraphics[width=1.0\linewidth]{{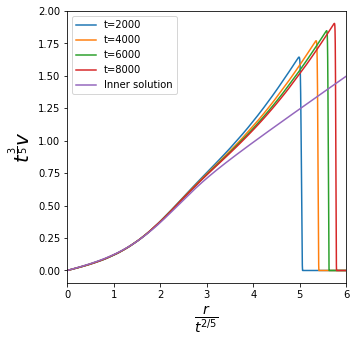}}
\caption{}
\end{subfigure}%
\begin{subfigure}{.35\textwidth}
\centering
\includegraphics[width=1.0\linewidth]{{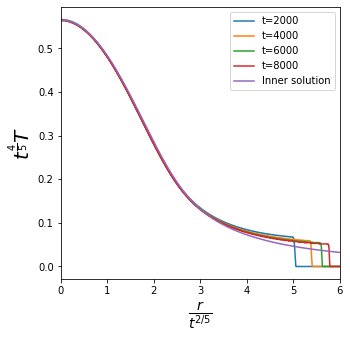}}
\caption{}
\end{subfigure}
\caption{{\bf NS solution for 2D hard point gas}: The three hydrodynamic fields, obtained from numerical solution of NS equations in Eq.~(\ref{eq:NS2d}) with ideal gas equation of state, are plotted in (a,b,c) according to the expected TvNS scaling form in Eqs.~(\ref{2DIGTvNS}) and in (d,e,f) according to the expected core scaling form [Eqs.~(\ref{eq:35Cscal})]. We can see that in (a,b,c) there is a very good collapse of data  everywhere except near the core. On the other hand in (d,e,f) we see excellent data collapse near the core. Thus the NS solutions exhibit the expected core and bulk scaling forms. In (a,b,c) we also see a good fit of the scaled data to the TvNS scaling functions $(G,V,Z)$, obtained from the exact solution [ Eqs.~(\ref{Z2deq},\ref{2DIG-TvNSsol}).
In (d,e,f) we see a reasonable fit in the core region to the core scaling functions $\tilde{G},\tilde{V},\tilde{Z}$, obtained from a numerical solution of Eqs.~(\ref{eq:38}) with boundary conditions obtained from the solution of the hard point NS equations (PDEs). 
The parameter values  were taken as $E_0=4.0$, $\rho_{\infty}=0.382$. 
\label{fig:2DIGNS}}
\end{figure}

\subsection{TvNS solution}
For this toy model, our first aim is to find the TvNS solution. To this end we take the following scaling ansatz for the form of the  hydrodynamic fields at long times:
\begin{subequations}
\begin{gather}
    \rho=\rho_{\infty}G(\xi),\\
    v=\frac{r}{t}V(\xi),\\
    T=\frac{r^2}{t^2}Z(\xi),
    \end{gather}
    \label{2DIGTvNS}
\end{subequations}
where $\xi={r}/{R}$, $R(t)=({E_0t^2}/{(A_2\rho_{\infty})})^{{1}/{4}}$ and $A_2$ is some dimensionless constant that we will determine later. Using these scaling ansatz, the Euler equations can be reduced to the following equations
\begin{subequations}
\label{eulerscaled2D}
\begin{gather}
    \Big(V-\frac{1}{2}\Big)G\xi\frac{dV}{d\xi}+\xi\frac{d}{d\xi}(ZG)-GV+GV^2+2GZ=0,\\
    \Big(V-\frac{1}{2}\Big)\xi\frac{dG}{d\xi}+\xi G\frac{dV}{d\xi}+2GV=0,\\
    -\Big(V-\frac{1}{2}\Big)\frac{\xi}{G}\frac{dG}{d\xi}+\Big(V-\frac{1}{2}\Big)\frac{\xi}{Z}\frac{dZ}{d\xi}+2(V-1)=0.
    \end{gather}
\end{subequations}
As a result of the scaling form we see that the total energy in the region $0<r<\xi R(t)$ is conserved. This leads to the following integral of motion~(see also Ref.~\cite{landau} for an alternate derivation):
\begin{equation}
    Z=\frac{V^2(1-2V)}{2(4V-1)}. \label{Z2deq}
\end{equation}
The Rankine-Hugoniot boundary conditions give us
\begin{subequations}
\begin{gather}
    G(1)=3,~~
    V(1)=\frac{1}{3},~~
    Z(1)=\frac{1}{18}.
    \end{gather}
\end{subequations}
These equations can be  solved to give
\begin{subequations}
\begin{align}
    \xi^4&=\frac{|4V-1|}{108V^2\Big(V-\frac{1}{2}\Big)^2}, \\
    G&=\frac{3\sqrt{3}\sqrt{|4V-1|}e^{-\frac{1}{2V-1}}}{e^3}.
    \end{align}
    \label{2DIG-TvNSsol}
\end{subequations}
Together with Eq.~(\ref{Z2deq}), these provide a complete solution of Eqs.~(\ref{eulerscaled2D}), except that the value of the constant $A_2$ is still undetermined. This can be found easily by using the conservation of energy, exactly as it was done in \cite{chakraborti,ganapa} for the case of 1D ideal gas:
\begin{equation}
A_2=    2\pi\int_0^{1}G(\xi)\Big(\frac{V^2(\xi)}{2}+Z(\xi)\Big)\xi^3d\xi.
\end{equation}
Plugging the exact solution for $G$, $V$, $Z$, we will obtain an expression for $A_2$ as an integral which cannot be evaluated exactly, but can be evaluated numerically. After doing the full calculation we obtain $A_2=0.3519359068$.

Using the exact solution, one finds that the following  $\xi\rightarrow 0$ behaviour of the various scaling functions:
\begin{align}
 \label{eq:50}
    V-\frac{1}{4}\sim\xi^4, ~~~
    G\sim\xi^2,~~~
    Z\sim \xi^{-4}.
\end{align}
\par

\par

\subsection{Core scaling solution}

Just like we observe a new scaling in the core region for hard discs, we expect a similar new scaling in the core region for our toy model. Following the same argument as in Sec.~(\ref{sec:2.2}), and using the fact that the $\xi \to 0$ behaviour of the TvNS solution, Eq.~(\ref{eq:50}), remains the same as for hard discs, we see immediately that again the size of the dissipative core  grows with time as $t^{2/5}$. It is clear also that the core scaling form will again be exactly as in Eq.~(\ref{eq:35Cscal}). Finally, we note that the higher order terms, in the virial expansion for pressure in the hard disc gas, do  not appear in the equations for the scaling functions $\tilde{G}(\eta),\tilde{V}(\eta),\tilde{Z}(\eta)$ in Eq.~(\ref{eq:38}). Hence we have the same ODE equations in the core as for our 2D  point particle gas. As before we need two boundary conditions to be determined either from NS data, or from the asymptotic matching condition, which is given by Eqs.~(\ref{eq:matching}). The boundary conditions are different for the hard point gas and will lead to a different solution for the scaling functions. 

\par
\subsection{Numerical results}
 In Fig.~(\ref{fig:2DIGNS}), we compare results from NS solutions with the predicted TvNS and core scaling forms. In Fig.~(\ref{fig:2DIGNS})(a,b,c), we see that the NS shock front matches with the TvNS prediction. We also see a clear TvNS scaling in the bulk region but significant departures in the core. Plotting the exact TvNS scaling  functions, we notice that they agree with the bulk NS results. In Fig.~(\ref{fig:2DIGNS})(d,e,f), we verify that the NS solutions satisfy the core scaling. The comparison between the numerical solution of the core scaling ODEs in Eqs.~(\ref{eq:38}) and that of the full NS equations is also good in the collapsed region. For the solution of Eqs.~(\ref{eq:38}), we took the boundary conditions from the full solution of the NS equations.
\par
\begin{figure}
\centering
\begin{subfigure}{.35\textwidth}
\centering
\includegraphics[width=1.0\linewidth]{{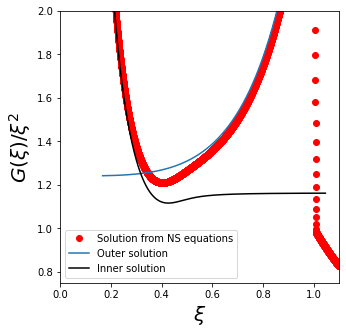}}
\caption{}
\end{subfigure}%
\begin{subfigure}{.35\textwidth}
\centering
\includegraphics[width=1.0\linewidth]{{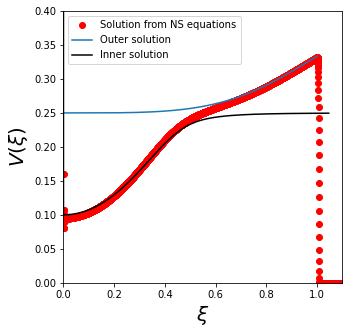}}
\caption{}
\end{subfigure}%
\begin{subfigure}{.35\textwidth}
\centering
\includegraphics[width=1.0\linewidth]{{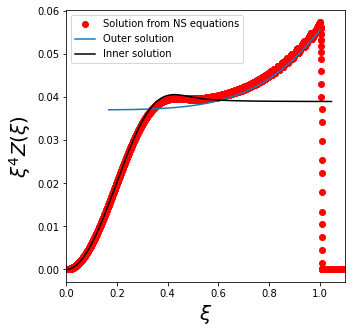}}
\caption{}
\end{subfigure}%
\caption{{\bf 2D hard point gas}: In this figure, the inner core scaling solution, the exact TvNS solution and the solution obtained from the full NS equations are plotted together. The boundary values for finding the inner solution were taken from the full solution of the NS equations. We see that the near core asymptotics of outer solution matches with the far core asymptotics of inner solution, which is nothing but the rule of asymptotic matching.\label{fig22}}
\end{figure}

In Fig.~(\ref{fig22})  we plot the exact TvNS solution, the core scaling functions and the solution from the full NS  equations.  We see a very good verification  of the rules of asymptotic matching. 

\section{Blast in  the hard rod gas}\label{sec:4}
In this section, we consider the blast problem in the 1D hard rod gas, and compare the predictions of MD simulations with those of NS equations and TvNS solution.
\par
We consider a gas of hard rods  of  length $a$. If all the hard rods are of equal masses, the particles simply exchange velocities during collisions, hence the system would be integrable and  never  reach  local equilibrium. Thus, as in \cite{chakraborti,ganapa}, we consider the alternate mass hard rod gas where successive rods on the line have masses $m$ and $M>m$ with mean mass $\Bar{m}={(m+M)}/2=1$. In units where $k_B=1$, the equation of state for this hard rod gas is given by:
\begin{equation}
    P=\frac{\rho T}{1-\rho a}.
\end{equation}
 
\subsection{TvNS solution}
The shock front in 1D grows with time as $R(t)=\Big(\frac{E_0t^2}{A_1\rho_\infty}\Big)^{\frac{1}{3}}$, where $\rho_\infty$ is the background density. As usual, we choose this growing length scale to define the TvNS scaling variable $\xi=x/R(t)$ and  the following scaling forms for the hydrodynamic fields:
\begin{align}
    \rho=\rho_\infty G(\xi),~~~
    v=\frac{x}{t}V(\xi),~~~
    T=\frac{x^2}{t^2}Z(\xi).
\end{align}
Plugging these into the Euler equations, we get the following ODEs for the scaling functions:
\begin{subequations}
    \label{1d-TvNSeq}
\begin{gather}
    -\frac{2}{3}\xi G'+\xi (GV)'+GV=0,\\
    -V+V^2-\frac{2\xi V'}{3}+\xi VV'+\frac{\xi ZG'}{G(1-\phi G)^2}+\frac{2Z}{1-\phi G}+\frac{\xi Z'}{1-\phi G}=0,\\
    -2Z-\frac{2\xi Z'}{3}+2ZV+\xi V Z'+\frac{2ZV}{1-\phi G}+\frac{2\xi ZV'}{1-\phi G}=0,
    \end{gather}
\end{subequations}
where $\phi=\rho_{\infty}a$ is the background packing fraction. 
As a result of the scaling form we see that the total energy in the region $0<r<\xi R(t)$ is conserved. This leads to the following integral of motion~\cite{landau}:
\begin{align}
\label{eq:exadd}
Z=\frac{V^2(1-3V/2)}{3V/2+3V(1-\phi G)^{-1}-1}.  
\end{align}
  The Rankine-Hugoniot conditions give the following boundary values needed to solve Eqs.~(\ref{1d-TvNSeq}):
\begin{align}
    G(1)=\frac{2}{1+\phi},~~~
    V(1)=\frac{1-\phi}{3}, ~~~
    Z(1)=\frac{(1-\phi)^2}{9}.
\end{align}
These equations have to be solved along with the equation which specifies the total energy, which will fix the constant $A_1$. For point particles, Eq.~(\ref{eq:exadd})  makes it possible to solve the TvNS equations in any dimensions. However, for the case of hard rods we have not been able to find a closed form solution. We have verified numerically that the numerical solution of the ODEs above have the same behaviour near the core as the TvNS solution of ideal gas obtained in \cite{chakraborti,ganapa}, i.e.,
\begin{align}
        G(\xi)\sim|\xi|^{1/2},~~
        V(\xi)=2/9,~~
        Z(\xi)\sim |\xi|^{-5/2},
\end{align}
for small $\xi$. This is expected as the small $\xi$ region is a region of low density making the virial corrections almost negligible and the gas ideal.
\par

\subsection{Core scaling solution}  Following Refs.~\cite{chakraborti,ganapa},  we take for the heat conductivity in NS to be of the form $\kappa=D_{\kappa}\rho^{1/3}T^{1/2}$. An analysis similar to the one we did in Sec.~(\ref{sec:2.2}) leads us then to a core growing as $X(t) \sim t^{38/93}$ and  the following  forms for the scaling near the core:
\begin{subequations}
\label{eq:35Cscal1D}
\begin{gather}
    \rho=t^{-\frac{4}{31}}\tilde{G}(\eta),\\
    v=t^{-\frac{55}{93}}\tilde{V}(\eta),\\
    T=t^{-\frac{50}{93}}\tilde{Z}(\eta),
    \end{gather}
\end{subequations}
where $\eta=xt^{-38/93}$.

\begin{figure}
\centering
\begin{subfigure}{.35\textwidth}
\centering
\includegraphics[width=1.0\linewidth]{{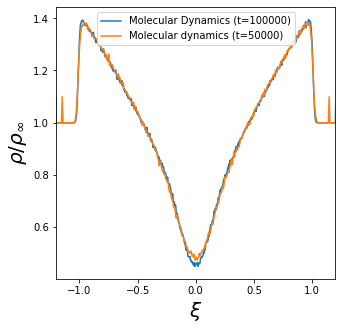}}
\caption{}
\end{subfigure}%
\begin{subfigure}{.35\textwidth}
\centering
\includegraphics[width=1.0\linewidth]{{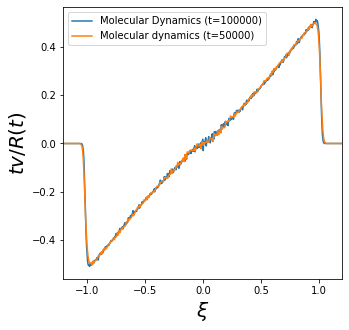}}
\caption{}
\end{subfigure}%
\begin{subfigure}{.35\textwidth}
\centering
\includegraphics[width=1.0\linewidth]{{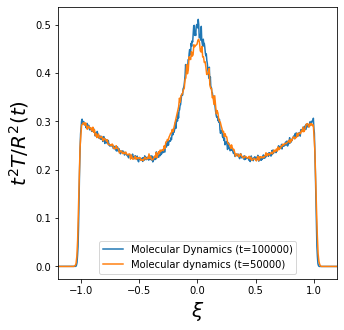}}
\caption{}
\end{subfigure}
\begin{subfigure}{.35\textwidth}
\centering
\includegraphics[width=1.0\linewidth]{{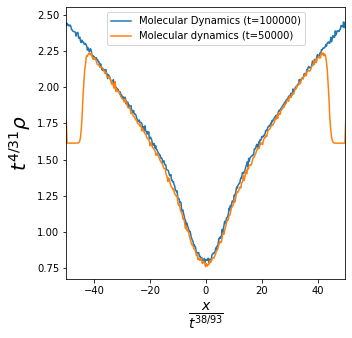}}
\caption{}
\end{subfigure}%
\begin{subfigure}{.35\textwidth}
\centering
\includegraphics[width=1.0\linewidth]{{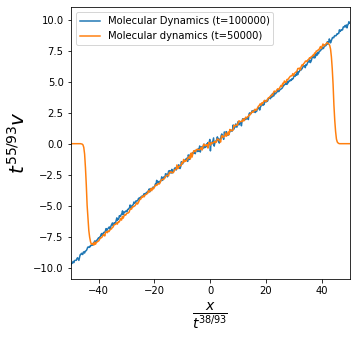}}
\caption{}
\end{subfigure}%
\begin{subfigure}{.35\textwidth}
\centering
\includegraphics[width=1.0\linewidth]{{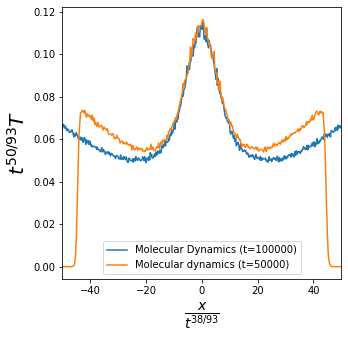}}
\caption{}
\end{subfigure}
\caption{ {\bf Hard rods}: Plot checking the TvNS scaling in (a,b,c) and the new scaling in the core (obtained by including the dissipative effects) in (d,e,f) from the results of MD simulation. We can see that there is a nice collapse of data at two different times in the bulk region in (a,b,c) and in the core region in (d,e,f). We have taken $\rho_{\infty}=0.4$, $E_0=0.4$.\label{fig29}}
\end{figure}
\begin{figure}
\centering
\begin{subfigure}{.35\textwidth}
\centering
\includegraphics[width=1.0\linewidth]{{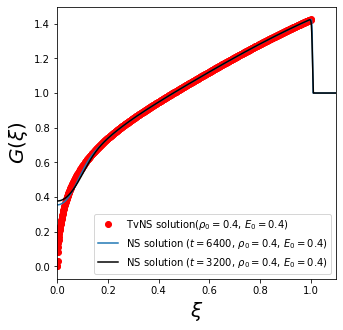}}
\caption{}
\end{subfigure}%
\begin{subfigure}{.35\textwidth}
\centering
\includegraphics[width=1.0\linewidth]{{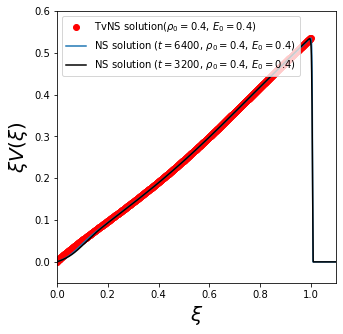}}
\caption{}
\end{subfigure}%
\begin{subfigure}{.35\textwidth}
\centering
\includegraphics[width=1.0\linewidth]{{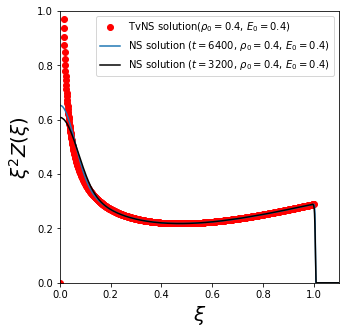}}
\caption{}
\end{subfigure}
\begin{subfigure}{.35\textwidth}
\centering
\includegraphics[width=1.0\linewidth]{{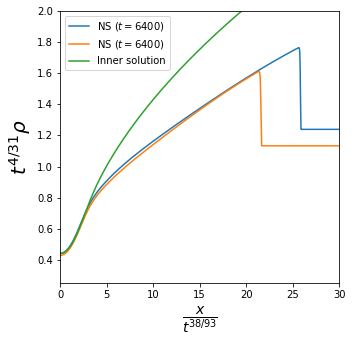}}
\caption{}
\end{subfigure}%
\begin{subfigure}{.35\textwidth}
\centering
\includegraphics[width=1.0\linewidth]{{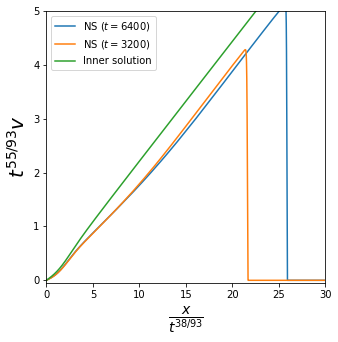}}
\caption{}
\end{subfigure}%
\begin{subfigure}{.35\textwidth}
\centering
\includegraphics[width=1.0\linewidth]{{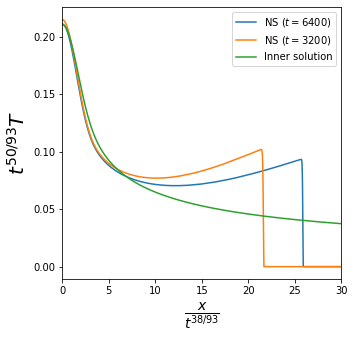}}
\caption{}
\end{subfigure}
\caption{ {\bf Hard rods}: Plot checking the TvNS scaling in (a,b,c) and the new scaling in the core (obtained by including the dissipative effects) in (d,e,f) from numerical solution of the NS equations. Also shown are the outer solution in (a,b,c) and the inner solution in (d,e,f). We can see that there is a nice collapse of data at two different times in the bulk region in (a,b,c) and in the core region in (d,e,f). We can also see that the outer solution matches with the NS data in the bulk region in (a,b,c) and the inner solution matches with the NS data in the core region in (d,e,f). We have taken $\rho_{\infty}=0.4$, $E_0=0.4$.\label{fig27}}
\end{figure}

We can solve for the core scaling functions similar to the way we did it for hard discs. Since the virial corrections to the equation of state are negligible in the core and the gas is almost ideal, we directly take the equations for the core scaling functions from \cite{ganapa}:
\begin{subequations}
\label{eq:45NSODEs1D}
\begin{gather}
    -\frac{4}{31}\tilde{G}-\frac{38}{93}\eta\tilde{G}'+(\tilde{G}\tilde{V})'=0,\\
    (\tilde{G}\tilde{Z})'=0,\\
    -\frac{25}{93}\tilde{G}\tilde{Z}-\frac{19}{93}\eta\tilde{G}\tilde{Z}'+\frac{\tilde{G}\tilde{V}\tilde{Z}'}{2}+\tilde{G}\tilde{Z}\tilde{V}'=(\tilde{G}^{1/3}\tilde{Z}^{1/2}\tilde{Z}')'.
    \end{gather}
\end{subequations}
As for the 2D case in Eqs.~(\ref{eq:38}) 
we observe again that  the change in the equation of state does not affect these equations  and only the thermal conductivity term makes an appearance in the core scaling equations.

\subsection{Numerical results} 
In our simulations, we have taken $m=4/5$ and $M=6/5$.
We verify the core scaling and bulk scaling from MD data in Figs.~(\ref{fig29}).
\par
We solved the TvNS ODEs numerically, and we solved the NS equations following the method of \cite{chakraborti,ganapa}. The plots comparing the two solutions are shown in Figs.~(\ref{fig27}a, \ref{fig27}b, \ref{fig27}c). We can see that the shock fronts predicted by the two solutions matches. Also there is a nice data collapse in the bulk for the NS solutions at two different times.
\par
We solve Eqs.~(\ref{eq:45NSODEs1D}) (with boundary conditions taken from the NS data) and compare the resulting solution with the full numerical solution of the NS equations in Figs.~(\ref{fig27}d, \ref{fig27}e, \ref{fig27}f). We see that there is agreement in the core, and that there is data collapse in the core of the NS solutions at two different times.
\par
\begin{figure}
\centering
\begin{subfigure}{.35\textwidth}
\centering
\includegraphics[width=1.0\linewidth]{{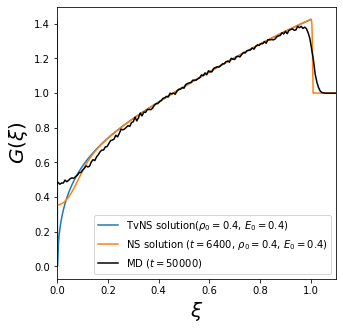}}
\caption{}
\end{subfigure}%
\begin{subfigure}{.35\textwidth}
\centering
\includegraphics[width=1.0\linewidth]{{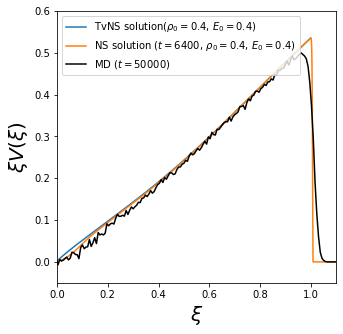}}
\caption{}
\end{subfigure}%
\begin{subfigure}{.35\textwidth}
\centering
\includegraphics[width=1.0\linewidth]{{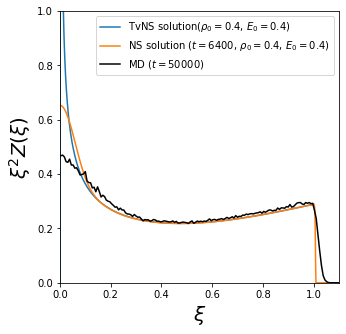}}
\caption{}
\end{subfigure}
\caption{{\bf Hard rods} Plot comparing the NS solution and TvNS solution with the MD simulations. We see that the shock front matches. We have taken $\rho_{\infty}=0.4$, $E_0=0.4$.\label{fig25}}
\end{figure}

 While solving the NS equations, we take (following \cite{chakraborti,ganapa}) heat conductivity $\kappa=D_{\kappa}\rho^{1/3}T^{1/2}$, with $D_\kappa=1$ and also a finite bulk viscosity $\zeta=D_{\zeta}T^{1/2}$, with $D_{\zeta}=1$ (setting this to zero does not change any of our conclusions).  
We also compared our NS numerical solution and TvNS solution with MD simulations (Fig.~(\ref{fig25})) and find that the shock front position agrees quite well with that obtained from MD simulations. Near the core, there are some discrepancies, possibly due to anomalous heat conduction in 1D, or due to large deviations from local equilibrium (also observed for the case for point particles in \cite{chakraborti,ganapa}).

\par

\section{Hard point gas in $d-$dimensions: Inner and outer solutions}
\label{sec:ideal}

In this section, we briefly discuss some features of the inner and outer solutions for  the point particle gas in arbitrary dimensions. 
As in Sec.~(\ref{sec:3}), we consider the limit of $\rho_\infty \to \infty$ and $a \to 0$ such that one has an ideal gas with finite transport coefficients. We first discuss the outer solution given by the TvNS-type analysis of the  Euler equations. Then we discuss the inner solution where one has to write the full NS equations including dissipation terms. 

\subsection{Outer solution}
\subsubsection{Euler  equations in $d-$dimensions}

From dimensional analysis we have in this case the shock front position $R(t)=\left(\frac{E_0t^2}{A_d\rho_{\infty}}\right)^{\frac{1}{d+2}}$. Behind the shock,  $0\leq r<R(t)$,  the radial velocity $v(r,t)$, density $\rho(r,t)$ and pressure $P(r,t)$ satisfy the Euler equations
\begin{subequations}
\begin{align}
\label{cont-eq}
&\partial_t \rho + \partial_r (\rho v) +\frac{d-1}{r}\,\rho v  = 0,\\
\label{entropy-eq}
&(\partial_t + v \partial_r)\ln\frac{P}{\rho^\gamma} =0,\\
\label{E-eq}
&\rho(\partial_t + v \partial_r) v +\partial_r P = 0,
\end{align}
\end{subequations}
where $\gamma$ is the adiabatic index.

The velocity of the shock wave is
\begin{equation}
\label{U-d}
U = \frac{dR}{dt} = \delta\,\frac{R}{t}, \qquad \delta\equiv \frac{2}{d+2}.
\end{equation}
The Rankine-Hugoniot conditions \cite{landau} describing the jump of the hydrodynamic variables on both sides of the shock wave read
\begin{equation}
\label{RH}
\frac{P(R)}{\rho_\infty U^2} = \frac{2}{\gamma+1}\,,~~
\frac{\rho(R)}{\rho_\infty}=  \frac{\gamma+1}{\gamma-1}\,, ~~
\frac{v(R)}{U}=\frac{2}{\gamma+1},
\end{equation}
in the case of the infinitely strong blast. 

\subsubsection{Self-similar solution}

Dimensional analysis assures that the hydrodynamic variables acquire a self-similar form 
\begin{equation}
\label{scaling}
v=\delta\,\frac{r}{t}\,V, \quad \rho=\rho_\infty\,G, \quad c^2=\delta^2\,\frac{r^2}{t^2}\,Z.
\end{equation}
Here $c^2=\gamma P/\rho=\gamma T$ is the square of the speed of sound and we have introduced constant factors of $\delta$ and $\delta^2$ in the definition of the scaling functions to simplify subsequent computations.  The dimensionless quantities $V, G, Z$ depend on the single variable
\begin{equation}
\label{xi-def}
\xi = \frac{r}{R}.
\end{equation}

One seeks the behavior of $V(\xi), G(\xi)$ and $Z(\xi)$ behind the shock wave, $0\leq \xi\leq 1$. The Rankine-Hugoniot conditions Eqs.~(\ref{RH}) become
\begin{subequations}
\begin{align}
\label{RH-V}
& V(1)=\frac{2}{\gamma+1},\\
\label{RH-G}
& G(1)= \frac{\gamma+1}{\gamma-1},\\
\label{RH-Z}
& Z(1)=\frac{2\gamma(\gamma-1)}{(\gamma+1)^2}.
\end{align}
\end{subequations}

Using energy conservation, one expresses $Z$ through the scaled velocity $V$ to give \cite{landau}:
\begin{equation}
\label{ZV}
Z = \frac{\gamma(\gamma-1)(1-V)V^2}{2(\gamma V -1)}.
\end{equation}

Plugging the ansatz Eqs.~ (\ref{scaling}--\ref{xi-def}) into Eq.~(\ref{cont-eq}) gives 
\begin{subequations}
\begin{equation}
\label{VG-eq}
\frac{dV}{d\ell}+(V-1)\,\frac{d\ln G}{d\ell}    = - d V,
\end{equation}
where $\ell=\ln\xi$. Similarly we transform Eq.~(\ref{entropy-eq}) into
\begin{equation}
\label{ZG-eq}
\frac{d\ln(Z/G^{\gamma-1})}{d\ell} = \frac{d+2-2V}{V-1}.
\end{equation}
\end{subequations}

Equations~ (\ref{VG-eq}--\ref{ZG-eq})  can be solved for arbitrary $d$ and $\gamma>1$. Having solved the problem, one can compute the energy
\begin{eqnarray*}
\label{energy}
E  &=& \int_0^R dr\,\,\Omega_d\, r^{d-1}\,\rho\left[\frac{v^2}{2}+\frac{c^2}{\gamma(\gamma-1)}\right] \nonumber \\
    &=& n_\infty\,\Omega_d\, \delta^2\,\frac{R^{d+2-a}}{t^2} \int_0^1 d\xi\,\xi^{d+1}\,\frac{(\gamma-1)V^3}{2(\gamma V -1)}\,G.
\end{eqnarray*}
Here $\Omega_d=2\pi^{d/2}/\Gamma(d/2)$ is the surface area of the unit sphere and we have used Eq.~(\ref{ZV}). Substituting Eq.~(\ref{R-d}) into the above expression for the energy we fix the dimensionless constant $A_d$:
\begin{equation}
\label{A-int}
A_d(\gamma) =\Omega_d\, \delta^2\,\frac{\gamma-1}{2} \int_0^1 d\xi\,\xi^{d+1}\,\frac{V^3}{\gamma V -1}\,G.
\end{equation}

In the following we limit ourselves to a monoatomic gas where $\gamma=1+2/d$. The adiabatic index generally depends only on the spatial dimension and independent of the interaction between particles \cite{fluid,Cercignani}. The monoatomic gases are relevant in astrophysics where even diatomic molecules are extremely rare. For the monoatomic gas, the Rankine-Hugoniot conditions Eqs.~(\ref{RH-V}--\ref{RH-Z}) become
\begin{equation}
\label{RH-d}
V(1)=\frac{d}{d+1}\,, \quad G(1)= d+1, \quad Z(1)=\frac{d+2}{(d+1)^2},
\end{equation}
 and Eq.~(\ref{ZV}) reduces to
\begin{equation}
\label{ZV:d}
Z = \frac{(1+2/d)(1-V)V^2}{(d+2)V -d}.
\end{equation}

Using Eq.~(\ref{VG-eq}) and Eq.~(\ref{ZG-eq}) with $\gamma=1+2/d$ and $Z$ given by Eq.~(\ref{ZV:d}) we express $dV/d\ell$ and $d(\ln G)/d\ell$ through the scaled velocity $V$:
\begin{subequations}
\begin{align}
\label{V-eq-d}
& \frac{dV}{d\ell}  = V\,\frac{X}{2D},\\
\label{G-eq-d}
 & \frac{d\ln G}{d\ell} = \frac{V}{V-1}\,\frac{Y}{2D},
\end{align}
\end{subequations}
where we  write in short
\begin{equation*}
\begin{split}
X & = [(d+2)V-d](d+2-4V), \\
D &= d-2(d+1)V+(d+1)(1+2/d) V^2,\\
Y &= (2 - d) d + (d-2) (2 + 3 d) V +  2 (d+2) (1- d) V^2.
\end{split}
\end{equation*}

Integrating Eq.~(\ref{V-eq-d}) yields an implicit solution for the scaled velocity
\begin{equation}
\label{V-d}
\xi  = c_d[(d+2)V-d]^\frac{2}{d^2+4}\, V^{-\frac{2}{d+2}}\,(d+2-4V)^{-\lambda_d},
\end{equation}
with
\begin{equation}
\label{c-lambda}
\begin{split}
c_d & = \left(\frac{d}{d+1}\right)^{\frac{2}{d+2}-\frac{2}{d^2+4}}\, \left(\frac{d^2-d+2}{d+1}\right)^{\lambda_d},\\
\lambda_d & = \frac{8 + 4 d + 10 d^2 - d^3 + d^4}{2 d (2 + d) (4 + d^2)}.
\end{split}
\end{equation}
We then divide Eq.~(\ref{G-eq-d}) by Eq.~(\ref{V-eq-d}) and integrate in $V$ to give 
\begin{equation}
\label{GV-d}
G  = C_d\,[(d+2)V-d]^\frac{d^2}{d^2+4}\, (1-V)^{-\frac{2d}{d-2}}\,(d+2-4V)^{\Lambda_d},
\end{equation}
with
\begin{equation}
\label{C-Lambda}
\begin{split}
          C_d & = (d+1)^{\frac{d^2}{d^2+4}-\frac{2d}{d-2}}\,\, d^{-\frac{d^2}{d^2+4}}\, \left(\frac{d^2-d+2}{d+1}\right)^{-\Lambda_d},\\
\Lambda_d & = \frac{8 + 4 d + 10 d^2 - d^3 + d^4}{2 (d-2) (4 + d^2)}.
\end{split}
\end{equation}
Finally,  $Z=Z(V)$ is given by Eq.~(\ref{ZV:d}). 
\subsubsection{Singular behavior}

Using Eq.~(\ref{V-d}) and Eq.~(\ref{GV-d}), one extracts the asymptotic behaviors near the center of the explosion ($\xi\ll 1$):
\begin{equation}
\label{GZ-diverge}
G  \sim \xi^\frac{d^2}{2}\,, \quad  Z^{-1}\sim V-\tfrac{d}{d+2}\sim \xi^\frac{d^2+4}{2}.
\end{equation}
In terms of the original coordinates
\begin{equation}
\label{r-T-diverge}
\rho  \sim r^\frac{d^2}{2}\,t^{-\frac{d^2}{d+2}}\,, \qquad  T \sim r^{-\frac{d^2}{2}}\,t^\frac{d(d-2)}{d+2}.
\end{equation}

This physically dubious behavior (the density vanishes, while the temperature diverges at $r=0$) indicates that the framework we have employed so far becomes incomplete near the center of the explosion. We rectify it by including dissipation.

\subsection{Inner solution}
\label{sec:NSF}
\subsubsection{Navier-Stokes equations}

Here we are considering the hard-point gas in the Boltzmann-Grad limit. Notice however that, even if 
we were to take the model of hard sphere  gas to ensure that there are collisions, our core analysis would not change. This would mean  virial corrections in the equation of state and corrections to the transport coefficients. However, since the core is a region of low density, these  corrections are negligible, and the core equations would be the  same as that of an ideal gas.
\par
The continuity equation is not affected by dissipative processes \cite{landau}. Thus for our radial flow Eq.~(\ref{cont-eq}) remains valid. The momentum equation is generally affected by viscosity---adding an extra term $\nabla \cdot \boldsymbol{\sigma}$ to the Euler equation yields the Navier-Stokes equation.  The viscous stress tensor is given by
\begin{equation*}
\boldsymbol{\sigma} = \mu  \left[\nabla{\bf v} +(\nabla{\bf v})^T-\tfrac{2}{d} (\nabla\cdot {\bf v}){\bf I}\right] +  \zeta (\nabla\cdot {\bf v}){\bf I},
\end{equation*}
where ${\bf I}$ is the unit tensor and the coefficients of shear and bulk viscosity are denoted by $\mu$ and $\zeta$.  Little is known about bulk viscosity. Fortunately, for dilute monoatomic gases the coefficient of bulk viscosity vanishes \cite{fluid}. Thus for the dilute hard sphere (HS) gas
\begin{equation}
\label{sigma}
\boldsymbol{\sigma} = \mu  \left[\nabla{\bf v} +(\nabla{\bf v})^T-\tfrac{2}{d} (\nabla\cdot {\bf v}){\bf I}\right]. 
\end{equation}
According to kinetic theory \cite{fluid,Cercignani}, transport coefficients exhibit similar behavior. For the dilute HS gas, these coefficients are proportional to $\sqrt{T}$. This is the classical prediction of kinetic theory.  More precisely, the coefficient of shear viscosity for the dilute hard sphere (HS) gas reads
\begin{equation}
\label{mu-HS}
\mu =  M_d\, a^{-(d-1)}\sqrt{mT},
\end{equation}
where $m$ and $a$ denote spheres' masses and  radii. The dimensionless amplitude $M_3$ is known only approximately \cite{fluid,Cercignani}. The coefficient  of thermal conductivity for the dilute HS gas is given by a similar formula
\begin{equation}
\label{kappa-HS}
\kappa =  K_d\, a^{-(d-1)}\sqrt{mT},
\end{equation}
which differs from the expression Eq.~\eqref{mu-HS} only by the dimensionless amplitude, $K_d$ instead of $M_d$.

The entropy equation is affected by heat conduction and viscous dissipation. To estimate the size of the core region where dissipative effects matter it suffices to keep heat conduction and ignore viscous dissipation. In this situation Eq.~(\ref{entropy-eq}) is replaced by equation
\begin{equation}
\label{entropy-heat}
 \rho T(\partial_t + v \partial_r)s =  r^{-(d-1)}\,\partial_r (r^{d-1}\kappa  \partial_r T),
\end{equation}
for the entropy per unit mass
\begin{equation}
\label{entropy-def}
s = \frac{d}{2}\,\ln\!\left(\frac{T}{\rho^{2/d}}\right).
\end{equation}

Using Eqs.~(\ref{entropy-def}--\ref{kappa-HS}) we recast Eq.~(\ref{entropy-heat}) into
\begin{equation}
\label{heat}
\rho^{1+2/d}\,(\partial_t + v \partial_r)\frac{T}{\rho^{2/d}} =  \frac{1}{r^{d-1}}\, \partial_r \big(r^{d-1} T^{1/2}\,\partial_r T\big). 
\end{equation}
We have set $K_da^{-(d-1)}\sqrt{m}=1$ in Eq.~\eqref{heat} to avoid cluttering the formulas; this factor can be always restored in the final results on dimensional grounds. 

\subsubsection{Heuristic analysis for core scaling}

Using \ref{heat} we deduce an estimate
\begin{equation}
\rho\,\frac{T}{t} \sim \frac{T^{3/2}}{r^2},
\end{equation}
which is combined with \ref{r-T-diverge} to give an estimate of the growth of the radius $X$ of the core where the heat transfer plays significant role
\begin{equation}
\label{size-d}
X  \propto R^{h_d}\,, \quad h_d=\frac{4 + 3d^2}{8 + 3 d^2}. 
\end{equation}

According to Eq.~(\ref{r-T-diverge}), the density vanishes and the temperature diverges at the center of the explosion. Heat conduction rectifies these predictions. Indeed, instead of using Eq.~(\ref{r-T-diverge}) at $r=0$, one should substitute $r=X$ into Eq.~(\ref{r-T-diverge}). This allows us to estimate the density, temperature and pressure in the core region including the center of the explosion
\begin{equation}
\label{rT:origin-d}
\rho^* \propto R^{-\nu_d}\,,      \quad T^*     \propto R^{-(d-\nu_d)}\,,      \quad p^*     \propto R^{-d},
\end{equation}
with $\nu_d = 2 d^2/(8 + 3 d^2)$.

\par

\section{Discussion}
\label{sec:discussion}
In this work, we presented the results of numerical solutions of the Navier-Stokes (NS) equations of dissipative hydrodynamics for a 2D hard disc gas, a 2D hard point gas, and a 1D hard rod gas. We revealed that the long-time solution has a double scaling form ---  it consists of  an outer solution described by the well-known TvNS scaling form, and an inner solution described by  different scaling functions. For the 2D gas, the inner solution is valid in the region $0<r \lesssim t^{2/5}$  while the outer solution is valid in the region $t^{2/5} \lesssim r < R(t)$, with $R(t)\sim t^{1/2}$ known quite precisely.  The inner solution arises due to the dominance of dissipative terms (namely, the viscosity and thermal conductivity terms in the NS equations) in the core of the blast. 
{

We also pointed out that molecular dynamics (MD) simulation data for both the 1D and 2D gases agree with the results from the solution of NS equations, except in the core. The MD data does satisfy the scaling forms corresponding to the inner and outer scaling solutions, however, the scaling  functions differ.
 This mismatch in the core occurs also for 1D point particles~\cite{chakraborti,ganapa}    and a possible reason could be that transport coefficients are anomalous in low dimensional systems. 
\par
Note that the ratio of the size of the core to the size of the blast $X(t)/R(t)$ in 1D (${t^{38/93}}/{t^{2/3}}\sim t^{-24/93}$ \cite{chakraborti,ganapa}) is much smaller than in 2D (${t^{2/5}}/{t^{1/2}} \sim t^{-1/10}$). } This ratio is even larger in 3D (${t^{62/175}}/{t^{2/5}}$). Despite the ratio being larger in 2D than in 1D, we find that dissipative corrections in the core do not affect the position of the shock front in 2D. In general $d$-dimensions~($d\geq2$), the  shock front position, $R(t)$, and   the core size, $X(t)$, grow as [see Sec.~(\ref{sec:ideal})  for the details of the derivation]:
\begin{equation}
    R(t) \sim t^{\frac{2}{(d+2)}},~~~~~X(t)\sim t^{\frac{2(4+3d^2)}{(d+2)(8+3d^2)}}.
\end{equation}
We present below a table showing the size of the shock front and the core, the hydrodynamic fields in the core and the ratio of the core energy ($E^*$) to the total energy in 2D and 3D [the details of the derivation are given in Sec.~(\ref{sec:ideal})].
\begin{center}
  \label{table:scaling}
\begin{tabular}{||c| c| c| c| c| c| c| c||} 
 \hline
 $d$ & $R$ & $X$ & $\rho^*$ & $T^*$ & $p^*$ & $v^*$ & $E^*/E_0$ \\ [0.5ex] 
 \hline\hline
 2 & ~$t^{1/2}$ ~& $t^{2/5}$ & $t^{-1/5}$  & $t^{-4/5}$ & $t^{-1}$ & $t^{-3/5}$ & $t^{-1/5}$\\
 \hline
 3 & ~ $t^{2/5}$ ~& ~$t^{62/175}$~ & $t^{-36/175}$ & $t^{-174/175}$ & $t^{-6/5}$ & $t^{-113/175}$ & $t^{-24/175}$ \\ [1ex]
\hline
\end{tabular}
\end{center}

The core is the region where deviations from local equilibrium are large (see also \cite{joy1}), and treating the deviation as a first order approximation (which is what is done in the NS framework) may not suffice. Apart from this, there is the issue of long time tails of correlation functions and the divergence of transport coefficients in dimensions $d\leq 2$~\cite{alder1970,beijeren,dhar2008,lepri2016}.  This can be  also be a reason for the mismatch between hydrodynamics and microscopic simulations observed in the core. Finally, 
another reason for the failure of hydrodynamics in the core could be the fact that the Knudsen number (which is the ratio of the mean free path to the length scale of variation of the hydrodynamic fields) in the core for the 2D case has a very slow decay than that in the bulk. In the core, the Knudsen number is $\sim {\rho^*}^{-1}t^{-2/5} \sim t^{-1/5}$, while that in the bulk is $\sim  \rho_{\infty}^{-1}t^{-1/2} \sim t^{-1/2}$. This reason is connected to the reason mentioned before, as the Knudsen number is a measure of the deviation from local equilibrium and how well the hydrodynamic limit is reached. 

Interesting open problems would be to resolve the observed disagreement between hydrodynamics and microscopic dynamics in the core for the blast problem, and to see if this disagreement disappears in higher dimensions.

\section*{Conflict of interest statement}
The authors declare that there is no conflict of interest.
\section*{Data Availability}
Data will be made available on reasonable request.
\section*{Acknowledgments}
We thank Amit Kumar and R. Rajesh for discussions and for permission to use data from Ref.~\cite{joy1}.  S.K.S and A. D. acknowledge support from the Department of Atomic Energy, Government of India, under project no. 19P1112R\&D. AD and PLK  acknowledge the support of the Erwin Schrodinger Institute where several discussions related to the work were held during the program `Large Deviations, Extremes and Anomalous Transport in Non-equilibrium Systems'. 
The numerical computations were done on ICTS clusters {\it{Contra}}, {\it{Tetris}} and {\it{Mario}}.

\bibliographystyle{unsrt}
\bibliography{references}



\end{document}